\documentclass[useAMS,usegraphicx]{mn2e}
\usepackage{amssymb}

\newcommand{\Dd}{D_{\rm d}}
\newcommand{\Ds}{D_{\rm s}}
\newcommand{\Dds}{D_{\rm ds}}
\newcommand{\Dp}{D_{\rm p}}

\begin{document}

\title{Statistics of gravitational lenses in the clumpy Universe}
\author[G. Covone et al.]
{G. Covone,$^{ 1}$ \thanks{E-mail: giovanni.covone@oamp.fr}
M. Sereno$^{2,3}$
and R. de Ritis\thanks{Deceased.}
\\$^1$ Laboratoire d'Astrophysique de Marseille,
traverse du Siphon, 13012 Marseille, France
%\\$^2$ INAF -- Telescopio Nazionale Galileo, Santa Cruz de La Palma, Spain
\\$^2$ Universit\`a degli Studi di Napoli ``Federico II,'' 
Via Cinthia, Complesso Universitario Monte S. Angelo, 
80126 Naples, Italy
\\$^3$ INAF-Osservatorio Astronomico di Capodimonte, Salita Moiariello, 16 
80131 Naples, Italy
}

\maketitle

\begin{abstract}

We evaluate the effect of small scale inhomogeneities  on
large scale observations within the statistics of gravitationally 
lensed quasars.
At this aim, we consider a cosmological model
whose large scale properties (dynamics, matter distribution) are the
same as in Friedmann-Lema\^{\i}tre models, but whose matter
distribution is locally inhomogeneous. 
We use the well known Dyer-Roder distances to allow a simple analytical
expression of the optical depth $\tau$, and pay particular attention on the
different role played by the different notions of distances 
(filled beam angular diameter distance and Dyer-Roder distances)
when calculating this quantity,
following the prescription from Ehlers \&  Schneider 
for a coherent formalism.
We find that 
the expected number of gravitationally lensed quasars
is a decreasing function of the clumpiness parameter $\alpha$.

\end{abstract}


\section{Introduction}

One of the major task in modern cosmology is the precise determination of the
parameters which characterize the assumed cosmological model.
In this {\em direct approach} (according to Ellis' (1995) terminology)
a theoretical description of the space-time
is postulated, the Friedmann-Lema\^{\i}tre-Robertson-Walker (FLRW) model, 
and its parameters are determined by fitting the observational data.

The corner stone of the observational support to the FLRW model is the 
existence and the high 
isotropy of the relic Cosmic Microwave Background Radiation
(CMWB).
Ehlers, Geren \& Sachs (1968) showed that 
if the background radiation appears to be {\em exactly} isotropic to a
given family of observers then the space-time is {\em exactly} FLRW.
Therefore, together with the Copernican Principle, 
we can prove the Universe to be FLRW
just from our own observations of the CMWB.
Stoeger, Maartens \& Ellis (1995) extended this result to the case of an almost
isotropic background radiation, which implies an almost FLRW space-time.
This important results is a firmer ground for the assumption of the 
FLRW model
to describe the large scale structure of the Universe, 
but it also makes clear that we need to understand the departures from 
a 
spatially  homogeneous model when interpreting observational data.
Indeed, 
departures form perfect homogeneity change the distance-redshift relation, and
this has to be taken into account when  fitting the FLRW parameters to
observations. 
However, cosmological observations are usually fitted just 
using relationships derived for homogeneous models.

In recent years, several authors have addressed this problem in the context of
the observations of distant Type Ia supernovae (e.g., Holz 1998, 
Holz \& Wald 1998, Kantowski, Kao \& Thomas 2000, Sereno et al. 2002,
Wang, Holz \& Munshi 2002, Pyne \& Birkinshaw 2004).
It has been shown that the noise due
to weak lensing magnification from small scale matter inhomogeneities
yields large errors on the luminosity measurement of high-$z$ supernovae.

In this paper we aim at investigating the possible
systematic errors
due to neglecting the effects of the local inhomogeneities 
in the distribution of
matter when evaluating the cosmological constant $\Lambda$ from
gravitational lenses statistics.
Statistics of gravitationally lensed 
multiply imaged quasars 
has been since a long time considered an useful 
tool to constrain the cosmological parameters, in particular
the cosmological constant \cite{turner,Fukugita92,Kochanek96},
and the properties of the lensing galaxies (e.g. Maoz \& Rix 1993,
Kochanek 1993).
Recently, Chae (2003) has shown that 
the observed gravitational lensing rate in the CLASS radio survey 
yields strong support to a flat cosmological model dominated by vacuum energy,
with
$\Omega_{\rm m} \simeq 0.3$ and $\Omega_\Lambda \simeq 0.7 $.
The precision of these results is limited 
at the moment by the uncertainty on
the knowledge of the luminosity function of the lensing galaxy population, their
density profile and their evolution since $z \simeq 1$ 
(e.g., Mao 1991, Chae 2003).
For instance, Cheng \& Krauss (2000) have shown that
constraints on cosmological parameter are strongly dependent on the choice of
galaxies parameters (see also discussion in Kochanek et al. (1999)).

Another major limit is the fact that only few systematic surveys
for multiply imaged quasars have been completed up to date
\cite{claeskens}; today in fact, the statistical uncertainties on
$\Omega_{\Lambda}$
are still dominated by the
Poisson errors from the small number of gravitationally lensed
quasars. 
In the near future, the most promising source for new lensed
quasars will be wide
field surveys (Kuhlen, Keeton \& Madau 2004) and
targeted followups of newly discovered quasars 
(Morgan 2002, Morgan et al. 2004).
For instance, the Sloan Digital Sky Survey\footnote{{\tt http://www.sdss.org/}}
will almost double the number of known gravitational lenses.
Next considerable increase in the number of gravitational lenses 
is expected by new telescopes like the
VLT Survey Telescope\footnote{{\tt http://www.na.astro.it/}},
which will allow very wide and deep optical surveys in the Southern
hemisphere. 
See also Kuhlen et al. (2004) 
for a discussion of other gravitational lenses surveys to become operational in
the next future.

While these large observational projects
will considerably improve the precision of the
results and the importance of the tool,
they also make necessary 
to consider more realistically the detail of the light propagation through the
observed not homogeneous universe.
As pointed out by Ellis (1995), small scale inhomogeneities in matter
distribution have a considerable effect on both observations
(Dyer \& Oattes 1988) and dynamics (Russ et al. 1997) at large scale.
Moreover, since 
the lensing effects of the small inhomogeneities on the propagation of light 
change the angular diameter distance-redshift relation 
(Schneider, Ehlers \& Falco 1992),
we focus here on this specific problem.
Many different approaches have been developed to study the
gravitational lensing in inhomogeneous cosmological models,
but the simplest one from an analytical point of view, and yet
efficient, is the one proposed by Dyer \& Roeder (1972, 1973), in
which the effect of the local inhomogeneities along the light
bundles are described by the so-called clumpiness parameter
$\alpha$ (see definition in next section).
In the following, we allow the clumpiness parameter to be a 
direction-dependent quantity,
which is function both of the line of sight to the source and its redshift 
(see, e.g. Wang (1999)).

The statistical properties of a sample of gravitational 
lenses include the frequency of multiply imaged quasars,
the distribution of the lenses and source redshifts, 
of the angular separation distribution and 
of the image multiplicity.
In this work we will focus on the discussion of the total
lensing probability, leaving a detailed
discussion of the other statistical properties
for a following paper.
Moreover, 
as we focus our attention on an effect which is independent from our present day
knowledge of the galaxy luminosity function and the dark matter
velocity dispersion, we do not consider these aspects in detail.

The paper is organized as follows. 
In Sect.~2 we define the cosmological model and 
discuss the relevant distances in the study of the
propagation of light.
In Sect.~3 we calculate the gravitational lenses rate, 
and in Sect.~4 we discuss its behavior as a
function of $\alpha$, considering different
gravitational lens models. 
Finally, systematic effects on the
estimate of the cosmological constant are discussed in Sect.~5,
and in Sect.~6 we sum up our results.

\section{Role of cosmological distances}

While cosmological models which are homogeneous at all scales are very
successful in describing the overall dynamics and evolution of the
Universe, they do not allow a detailed description of the lensing
phenomena. As a fact, all the gravitational lensing phenomena (bending of light
rays, deformation of images, and formation of multiple images) are only
possible in a clumpy universe, (see, e.g., 
the discussion by Krasi\'nsky \shortcite{kra97}).
Therefore, 
a coherent approach needs an inhomogeneous model.

However,
in the statistical analysis of gravitational lenses, as
well as in analysis of other astronomical observations, perfect homogeneity
is often assumed (see, e.g., discussion in Buchert (2000) 
about this point).
An important reason for this choice is the fact that in
homogeneous space-times we have simple relations between the proper distance
and the cosmological distances, i.e. the luminosity distance and the
angular diameter distance (e.g., Kayser, Helbig \& Schramm
1997). In inhomogeneous cosmological models, these
relations are much more complicated, and Friedmann-Lema\^{\i}tre (FL) distances
are not generally a good approximation to be used in the determination of
cosmological parameters from a given set of observational data.

The relevant distance in gravitational lensing, the angular diameter
distance $D$, is operationally defined by the square root of the ratio of
the area $dA$ of a celestial body to the solid angle $d \Omega$ it subtends
at the observer \shortcite{SEF}:

\begin{equation}
D  \equiv \sqrt{\frac{dA}{d \Omega}}.
\label{distance-def}
\end{equation}
In a homogeneous universe, the angular diameter distance 
can be derived from the proper distance
$\Dp$ using the following relation
\begin{equation}
D (z) = \frac{\Dp (z) }{1+z}.
\end{equation}
This relation does not hold anymore in a clumpy universe. The basic reason
lies in the fact that the proper distance is related to the global geometry
of the universe, while the relation between the angular diameter distance
and the redshift is determined mainly by the local matter distribution.

In this paper we
assume a cosmological model  which is locally inhomogeneous, but
homogeneous at very large scale, according to some density averaging rule
(see, e.g., Krasi\'nski (1997)).
In other words, we assume that
the overall dynamics does not differ from the dynamics of a
homogeneous cosmological model.
This approximation is well justified: 
by averaging the Friedmann equation, Russ et al. (1997) showed
that the influence of small scale
clumpiness on the global expansion factor is negligible
(so, for instance, the age of the Universe can be evaluated using the FLRW
relation with the Hubble constant, with negligible errors), 
while the
distance-redshift relation is significantly affected
(so that the measurements of the Hubble constant via 
measurements of standard candles magnitudes have not negligible systematic
errors).

In order to describe the degree of inhomogeneity in the 
local distribution of matter, 
we use a generalized notion of the so-called
clumpiness parameter 
(e.g., Dyer \& Roeder 1972, Schneider et al. 1992), introduced by Wang (1999).
The clumpiness parameter $\alpha$ was first introduced by 
Dyer \& Roeder (1972, 1973, hereafter DR), 
when writing a differential equation for the angular diameter distance in 
locally clumpy cosmological models,
and it was defined as the fraction of homogeneously dstributed matter within 
a given light cone.

Starting from the equations for scalar optics (e.g., Zeldovich
\shortcite{Zeldovich}, Kristian \& Sachs \shortcite{sachs}),
DR derived a second order
differential equation for the diameter of the light ray bundle propagating
far away from any clumps (i.e. in regions where $\alpha < 1$), assuming a
negligible shear. For a pressure-less universe with cosmological constant
$\Lambda$, the DR equation reads:
\begin{eqnarray}
\nonumber
(1 &+&  z) \left \{ \Omega_{\rm m} (1+z)^3  + \Omega_K (1+z)^2 + 
\Omega_{\Lambda} \right \} \frac{d^2 D}{d z^2}  +
\\ \nonumber
&& \left \{ \frac{7}{2} \,  \Omega_{\rm m}  \, (1+z)^3 + 3 \, \Omega_K \,
(1+z)^2 + 2
\Omega_{\Lambda} \right \} \frac{d D}{d z} 
\\ 
%\nonumber
&& + \frac{3}{2} \, \alpha \, \Omega_{\rm m} (1+z)^2 D = 0 \, ,
\label{eqn:DR}
\end{eqnarray}
with initial conditions $D(0)=0$ and $D'(0) = c/H_0$.
Here  $\Omega_K
\equiv 1 - \Omega_{\rm m} - \Omega_{\Lambda}$, 
since we neglect the contribution
from any relativistic fluid or radiation.
For $\alpha=1$ (filled beam case) we recover the angular diameter distance,
while for $\alpha=0$
we have the well-known empty beam approximation.
For a detailed discussion of
the solutions of the equation (\ref{eqn:DR}) within 
quintessence cosmological models we refer to Sereno et al. (2000). 
For the following, 
It is useful to introduce the dimensionless distance $r$:
\begin{equation}
r(z,\Omega_{\rm m},
\Omega_{\Lambda}, \alpha) \equiv \, \frac{H_0}{c} \,  D (z, H_0, \Omega_{\rm m},
\Omega_{\Lambda}, \alpha) \, ,
\end{equation}
and the symbol $r_1$ for the dimensionless angular diameter distance in
the filled beam case. 

Note, however that the DR equation (\ref{eqn:DR}) is well defined for any $\alpha
> 0 $,
and in its derivation the mass density is never required to be uniform.
This allows to consider the clumpiness parameert as a local variable,
as done in Wang (1999) to descibe the weak lensing magnification of
distant standard candles.
Therefore, in the following we assume the clumpiness parameter to be a function
both of the line of sight and the redshift.
Given a source at resdhift $z$ in a specific inhomogeneous cosmological model, 
for any line of sight to the observer the clumpiness parameter $\alpha$ 
is calculated via equation (\ref{eqn:DR}), where the distance $D$ is given 
by numerical simulations.
As a consequence, a complete description of the light propagation in the clumpy
Universe needs the knowledge of the probablity distribution function
(PDF) for values of the clumpiness parameter. 
We return to this point in Sect.~5

\begin{figure}
\centerline{\scalebox{0.45}{\includegraphics{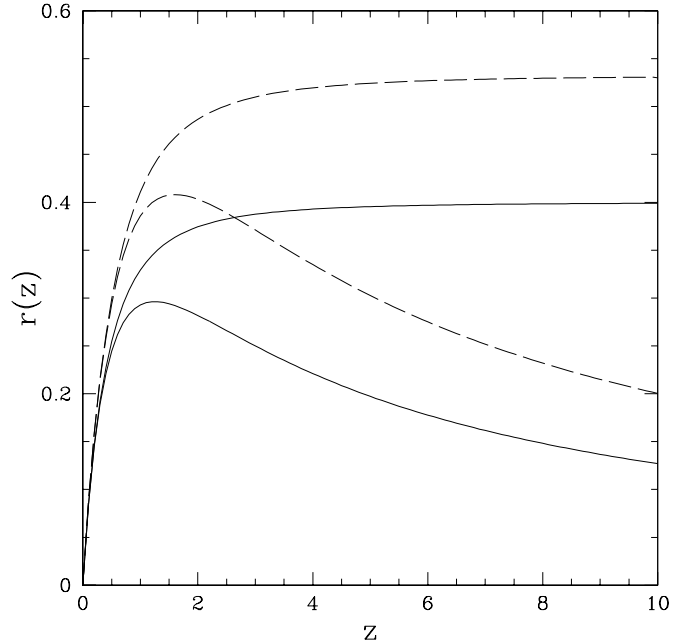}}}
\caption{The Dyer-Roeder distance $r(z)$ for two values of the
cosmological constant ($\Omega_{\Lambda} = 0.0$ (continuous curves),
$0.7$ (dashed curves)) in the empty beam case (upper curves)
and in the filled beam case (lower curves). Space-time is flat.}
\label{fig:DRdistance}
\end{figure}

\begin{figure}
\centerline{\scalebox{0.45}{\includegraphics{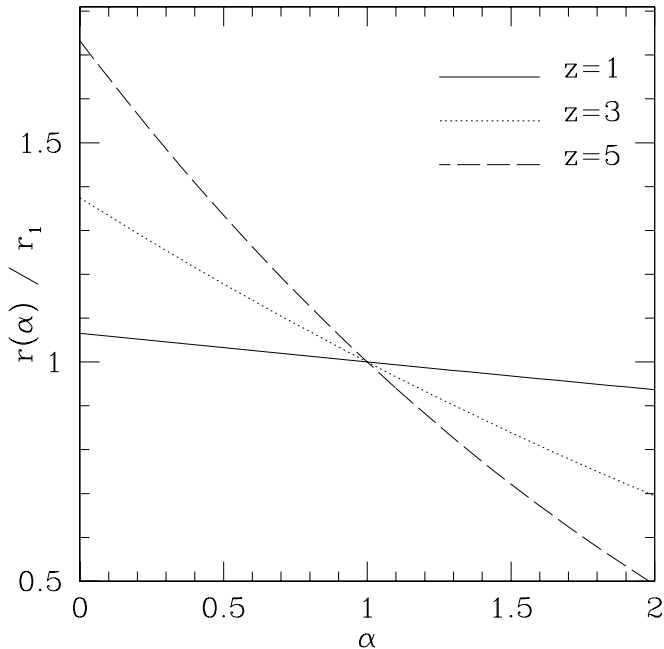}}}
\caption{DR distance versus the clumpiness parameter,
relatively to the filled beam case
at several value of the redshift. $\Omega_{\Lambda}=0.7$ in a flat
space-time.}
\label{Fig:figura2}
\end{figure}

Let us now consider in more detail the effect of inhomogeneities in observations 
meant to measure the cosmological constant. 
The DR distance $r$ is a strongly
decreasing function of $\alpha$, at fixed redshift (Schneider et al. 1992): 
therefore, a larger fraction of matter in clumps
partially masks the effect of a larger cosmological constant
when evaluating angular diameter and luminosity distances
(Fig.~\ref{fig:DRdistance}),
since there a smaller contribution from the Ricci convergence.
On the other side, along light beams characterized by  $\alpha > 1$ 
(i.e., propagating in overdense regions)
agular diameters distances are lower than in the
filled beam case.
For this reason a large amount of local clumpiness can result 
in a lower value for
$\Lambda$ when fitting the observational data.
In Fig.~\ref{Fig:figura2} we show the ratio of the DR distance 
for the empty beam case relatively to
the filled beam for a few value of the source redshifts.
Up to redshift $ \sim 1$,
the differences are not important in the 
DR distance itself 
(although, they maybe not  negligible for several astronomical observable 
quantities).
The role of the local clumps becomes more and more important at higher
$z$.

Though the assumed cosmological model is not the most satisfactory
to describe inhomogeneities, 
since it has not a firm theoretical
basis in the framework of General Relativity (i.e., it is not a
solution of the Einstein field equations), it allows a simple and
efficient description of the light propagation through a clumpy
universe. This model has been discussed in detail in Schneider et al. (1992)
and Seitz, Schneider \& Ehlers \shortcite{seitz}.
Ehlers \& Schneider \shortcite{ES} have introduced a
self-consistent formalism to study the gravitational lensing in a
clumpy Universe. In particular, they stressed
the different roles that the different notions of cosmological
distances have in this model. 
Namely, when we consider quantities
which are related to the global geometry of the assumed
cosmological model (such as volumes), it is necessary to use the
FL angular diameter distance. Then, the volume element (i.e. the
volume of a shell with proper thickness $dl$) is
\begin{equation}
dV = 4 \, \pi \, \left( \frac{c}{H_0} \right)^2 \, r_1^2 \, dl \, .
\label{eq:volume}
\end{equation}
In other words, the volume element
$dV$ does not depend on the local inhomogeneity degree. 
This is consistent with the fact that in the locally inhomogeneous
model, on large angular scales (i.e., larger than 
$\theta \sim 10''$, see Linder (1988)), the
distance-redshift relation is the one computed in the FLRW models, for any
source redshift.
This is also in agreement with the results from 
N-body numerical simulations in Cold Dark Matter scenarios 
(Tomita 1998), where it has been
shown that the dispersion in 
values of $\alpha$ along the different light paths becomes increasingly
larger as the angular is as small as a few arcsec.
%}
We also note that 
equation (\ref{eq:volume}) can be read as the definition of the
angular diameter distance in homogeneous models (Schneider et al.
1992).

On the other side, 
there a simple reason why,
when considering gravitational
lensing phenomena, the DR distance has to be used:
light deflection modifies the cross area of the light bundle
and the Ricci focusing term is a linear
function of the amount of matter within the ray bundle 
(see, e.g., Schneider et al. 1992, sect.~3.4).
As a consequence, all the
distance dependent quantities which play a role in the description of the
lensing phenomena are functions of the DR distance,
and, in particular, the strong lensing cross section $\sigma$.

Let us consider in more detail the cosmological strong lensing probability
$\tau$.
This is defined as the probability that a
light source at redshift $z_s$ is multiply imaged by a deflector
at $z<z_s$. In the expression of $\tau (z_s)$ there are two
different physical quantities which are functions of combinations
of distances: the cross section for multiple images $\sigma$, and
the volume element $dV$.

The cross section $\sigma$ is defined as the area in the lens
plane for which multiple imaging occurs for sources behind it.
This quantity depends on the redshift $z$ of the lens and a set
$\chi$ of astrophysical parameters which characterize the
gravitational lens model.
For the most generally used models (point-like mass distributions
and isothermal spheres) the cross section depends on a
particular combination of distances:

\begin{equation}
\label{sigma-def}
\sigma (z, z_s, \chi) = f \left ( \frac{\Dds \Dd}{\Ds}, \chi   \right) \, ,
\end{equation}
where $\Ds, \Dd $ and $\Dds $ are the DR angular diameter
distances between the observer and the source, the observer and the
lens and between the lens and the source, respectively.
We remark that the overall effect 
of the clumpy distribution of matter along
the light rays 
on the lensing probability
is not due to any change of the volume element $dV$, see
equation (\ref{eq:volume}), but it is only because of the 
dependence of the strong
lensing cross section $\sigma$ on the clumpiness parameter.

In the following we will consider the projection of the cross section on
the source plane (located at redshift $z_s$):
\begin{equation}
{\hat \sigma} (z, z_s, \chi) = \left[ \frac{r (z_s,\alpha)}{r(z,\alpha)}
\right]^2 \sigma (z, z_s, \chi),
\label{sigma-projected}
\end{equation}
where the DR distances are considered. 
This quantity allows a more clear and
compact definition of the lensing probability and it is the natural
quantity to consider in the assumed cosmological model \cite{ES}.
It is important to note that the quantity
${\hat \sigma}$ is not in general a function of the 
the distance combination $\Dds \Dd / \Ds $: as a consequence, 
the point-like mass and isothermal sphere 
cross sections are functions of different distances combinations (see Sect. 4).
We now calculate the probability of strong
lensing phenomena and evaluate its dependence on $\alpha$, and then 
determine explicitly the distance
functions for these two gravitational lens models.

\section{The statistics of gravitational lenses}

In this section, we derive the formulae for the statistics of
gravitational lensing, following mainly the formalism discussed in Ehlers
\& Schneider \shortcite{ES}, considering in particular the proper role of 
the two types
of distances.
Let us consider a statistical ensemble of cosmological sources at redshift
$z_s$ and a set of comoving gravitational lenses with number density
$n(z)$. If we neglect gravitational lenses evolution and mergings, the
comoving number density of lenses is conserved: $n(z) = n_0 (1+z)^3$, where
$n_0$ is the local density. The number of gravitational lenses in a shell
with volume $dV$ is then
\begin{eqnarray}
N (z) &=& n(z) dV \nonumber \\
&=& n(z) A(z) \frac{dl}{dz} dz,
\end{eqnarray}
where $A(z)$ is the area of the sphere located at redshift $z$.

The probability $d \tau (z, z_s, \chi)$ that a quasar at $z_s$ is multiply
imaged by a gravitational lens in the redshift range $(z, z+dz)$ is defined
as the fraction of the area of the sphere at $z = z_s$ (i.e., the fraction
of the sky) covered by the gravitational lenses cross sections ${\hat
\sigma} (z,z_s,\chi)$. This definition implicitly assumes that the
projected cross sections do not overlap,
which is
equivalent to state that $d \tau (z, z_s, \chi) \ll 1$.
The area covered by the projected cross sections of the gravitational
lenses in $(z, z+dz)$ is therefore

\begin{equation}
n(z) {\hat \sigma} (z,z_s, \chi) \, A(z)
\frac{ dl}{dz} \, dz \, .
\end{equation}

According to the definition, the differential lensing probability then reads
\begin{eqnarray}
\label{eq-01}
d \tau (z, z_s) &=& n(z) {\hat \sigma} (z,z_s, \chi) \frac{A(z)}{A (z_s)}
\frac{dl}{dz} dz \nonumber \\ 
%\nonumber &=& n(z) \sigma (z, \chi)
%\left [\frac{r(z_s, \alpha)}{r(z, \alpha)} \right]^2
%\frac{A(z)}{A (z_s)}
%\frac{dl}{dz} \, dz \\
&=& n(z) \sigma(z, \chi) \left [\frac{r(z_s, \alpha)}{r(z, \alpha)} \right]^2
\left [\frac{r_1(z)}{r_1(z_s)} \right]^2 \frac{dl}{dz} dz.
\end{eqnarray}

In the particular case $\alpha = 1$, the distance ratios factorize, so that
we obtain the more common expression
\begin{equation}
d \tau (z, z_s) = n(z) \sigma (z, z_s, \chi) \, \frac{dl}{dz} \, dz \, ,
\label{d-tau-1}
\end{equation}
i.e., the definition of differential lensing probability in cosmologies
which are homogeneous at all scales. 
Therefore, in this case, the relevant distance
combination is exactly the ratio $\Dds \Dd / \Ds $ which appears in
equation (\ref{sigma-def}), and does not depend on the particular choice of
the gravitational lens model.

Let us now evaluate the explicit expressions for the lensing probabilities
in equation (\ref{d-tau-1}). 
The quantity $dl$ can be written in the following way:
\begin{equation}
dl = -c \,  dt = c \frac{1}{H(z)} \frac{dz}{1+z} \, ,
\end{equation}
where we considered the past light cone, and $H(z)\equiv {\dot{a}}/ a$, $ a
= 1/ (1+z) $ being the  normalized cosmological scale factor. In a FLRW
cosmological model
\begin{equation}
H(z) = H_0 \sqrt{\Omega_{\rm m} (1+z)^3 + 
(1-\Omega_{\rm m} - \Omega_{\Lambda}) (1+z)^2
+ \Omega_{\Lambda}} \, .
\end{equation}

Hereafter we focus our attention to flat cosmological models,
i.e., with $\Omega_{\Lambda} + \Omega_{\rm m} = 1$, 
as these are preferred by inflationary
scenarios and strongly supported by many 
recent observational evidences \cite{wang}. 
Then, the differential lensing probability reads
\begin{eqnarray}
\label{d-tau-2}
d \tau (z, z_s) &=& n_0 \frac{c}{H_0} \sigma(z, z_s, \chi) \left
[\frac{r(z_s)}{r(z)}
\right]^2 \left [\frac{r_1(z)}{r_1(z_s)} \right]^2 \times 
\nonumber \\
& &
\frac{(1+z)^2}{\sqrt{\Omega_{\rm m} (1+z)^3 + \Omega_{\Lambda}}} \, dz \, .
\end{eqnarray}
It is evident that the properties of the functions $d \tau$  (and
$\tau$) with respect to $\alpha$ are determined only by the strong
lensing cross section $\hat \sigma (z, z_s)$. 
Thereby, it is
necessary to consider only those functions of the DR distances
which enter their expressions.
In other words,
the functions $\tau $ and ${\hat \sigma}$ have the same
qualitative behaviors with respect to the clumpiness parameter.
Then, our next point is to investigate the behavior of the quantity
${\hat{\sigma}} (\alpha)$.

\section{Properties of DR distances combinations}

In this section we analyze in detail the cross sections of some
general models of gravitational lenses and the involved
combinations of  DR distances, in order to evaluate qualitatively
the dependence of the lensing probability $\tau$ on the clumpiness
parameter. Asada \shortcite{Asada98} has investigated the
analytical properties of several combinations of DR distances as
functions of the clumpiness parameter, and he also deduced some
consequences on the observable quantities: the deflection angle,
the time delay, and the lensing probability. 
Here we focus on the study of the quantities which directly enter the 
optical depth.
We consider
the following models: a point mass distribution, singular
isothermal sphere, and isothermal sphere with a non zero core
radius. In this section we only consider
the DR distances.

\subsection{Point-like gravitational lens}

\begin{figure}
\centerline{\scalebox{0.45}{\includegraphics{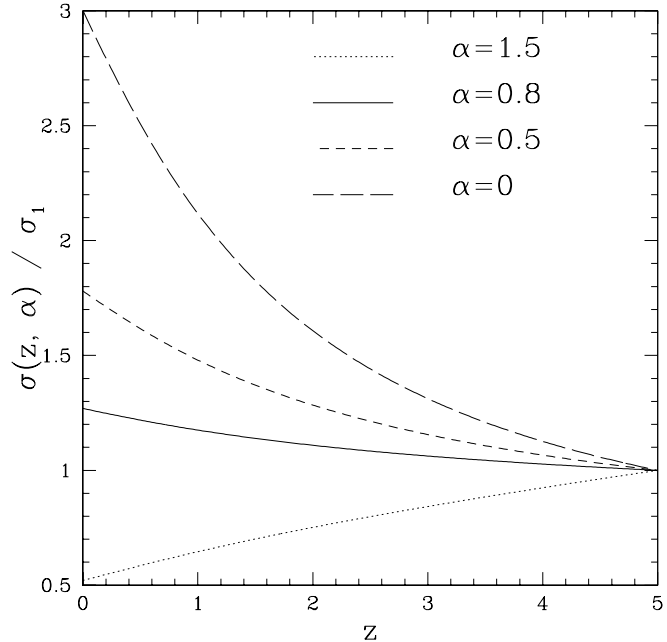}}}
\caption{The projected cross sections of a point-like
gravitational lens for $\alpha = 0, 0.5, 0.8$, normalized to filled beam
case ($\alpha=1$), versus the lens redshift. 
Source is at $z_s = 5$, and cosmological parameters
are $\Omega_{\Lambda}=0.7$ and $\Omega_{\rm m}=0.3$}
\label{fig-lentepunto}
\end{figure}

A point-like gravitational lens produces two images, whatever the source
position is; so, strictly speaking, the given definition of strong lensing
cross section does not apply here. Anyway, it is natural to assume such a
cross section ({\em on the lens plane}) to be the disk with radius equal to
Einstein radius $r_{\rm E}$. So, the cross section {\em on the source plane}
reads:
\begin{eqnarray}
\nonumber
{\hat \sigma} &=& \left( \frac{\Ds}{\Dd} \right) ^2 \pi r_{\rm E}^2 \\ &=&
\frac{4 \pi G M}{c^2} \frac{\Ds \Dds} {\Dd} \, ,
\end{eqnarray}
where $M$ is the mass of the lens.
This quantity
is a decreasing function of the clumpiness parameter
$\alpha$, as it is shown in
Fig.~\ref{fig-lentepunto}.
Consequently, at smaller value of the clumpiness parameter, 
the lensing probability is higher, when 
considering the lensing cross
section for a point-like distribution of mass.

Consider that the distances ratio $\Dd \Dds / \Ds$ in the expression of
the Einstein radius $r_{\rm E}$ 
is a decreasing function of the clumpiness parameter
\cite{Asada98}, but it is not the
right function of DR distances which enters the expression of the lensing
probability in a DR cosmological model, as shown above. As a fact,
considering this ratio is not coherent with the assumptions on the
cosmological model \cite{ES}, and would wrongly lead to predict that the
lensing probability decreases for decreasing values of the clumpiness
parameter, as in Asada \shortcite{Asada98}.

\subsection{Isothermal spheres}

\begin{figure*}
\centerline{\scalebox{0.4}{\includegraphics{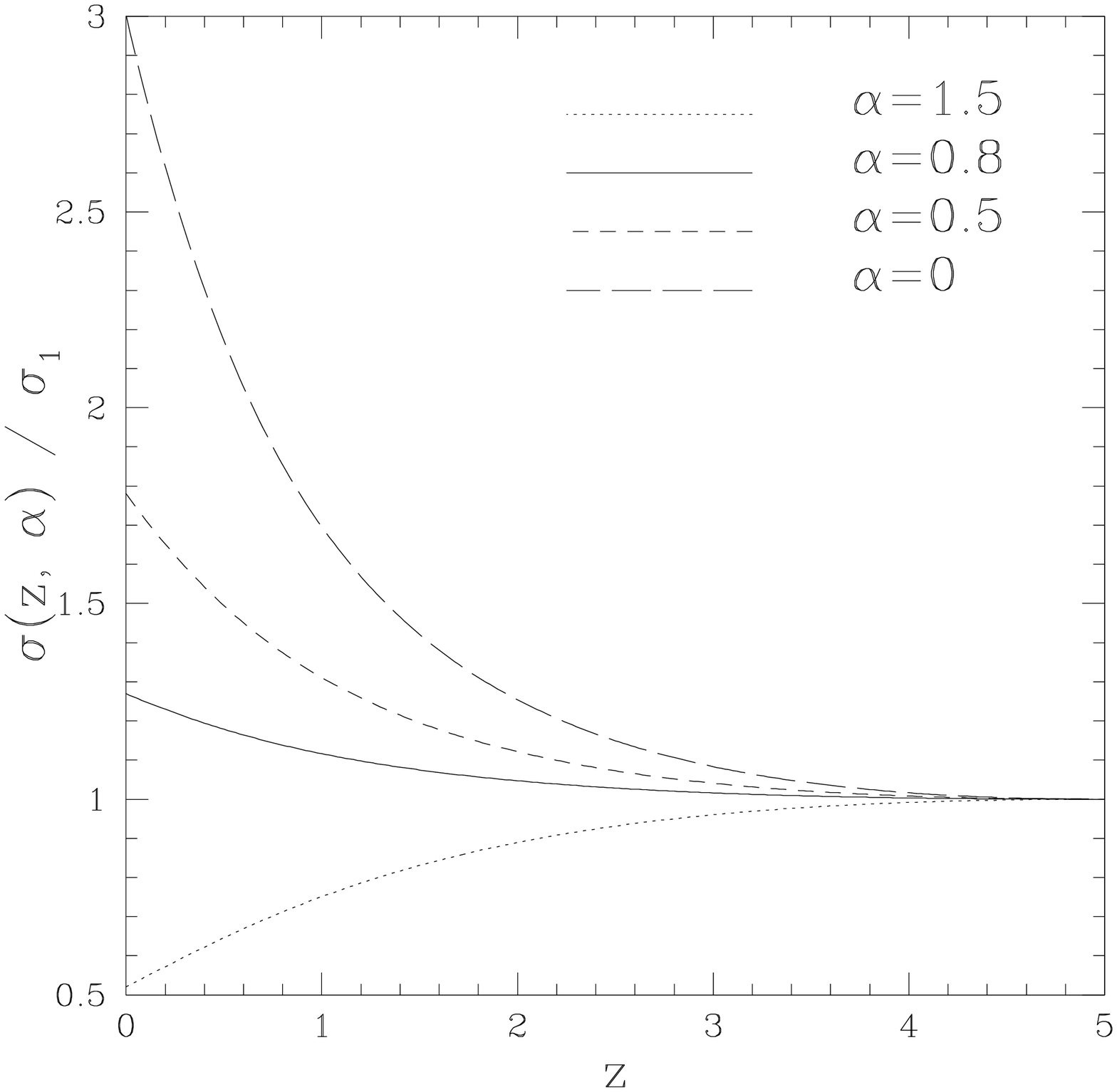}
\includegraphics{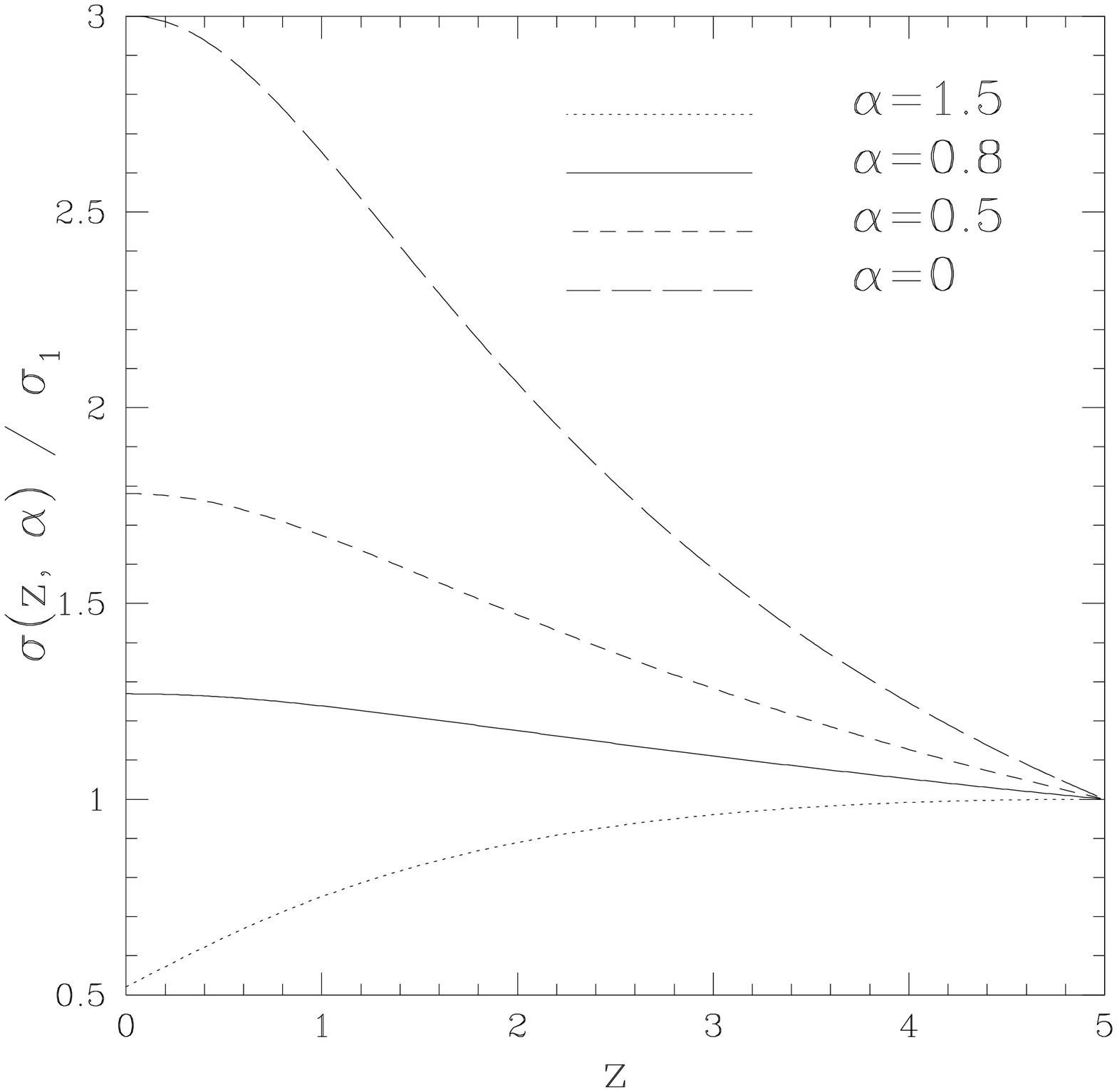}}}
\caption{Projected cross sections relatively to the filled beam
case for a singular isothermal sphere (left-hand panel) 
and an isothermal sphere with
core radius $\xi_{\rm c} = 10$ pc (right-hand panel), versus the lens redshift.
The source is located at $z_{\rm s}=5$ and
cosmological parameters are
$\Omega_{\Lambda}=0.7$, $\Omega_m=0.3$. 
% and several values of the clumpiness parameter.
For lens redshift $z \ll 1$, the two lens models
have the same limit.
The same qualitative results
holds for any value of $z_s$ and $\Omega_{\Lambda}$.}
\label{fig-IS}
\end{figure*}

Let us consider now the gravitational lenses to be isothermal
spheres, and let us study the dependence of the cross section on
$\alpha$. 
Many authors have shown that singular isothermal spheres (SIS) allow a detailed
description of the matter distribution in individual gravitational lenses 
(see, e.g., Rusin et al. 2002), and are therefore used in statistical 
analysis of the lensing galaxies population (e.g., Chae 2003).
SIS are characterized by the surface mass density

\begin{equation}
\Sigma({\mathbf \xi}) = \frac{\sigma^2_v}{2 G} \frac{1}{\xi},
\end{equation}
where ${\mathbf \xi}$ is the position vector on the lens plane, and
$\sigma_v$ the line-of-sight velocity dispersion. See, e.g., Hinshaw
\& Krauss \shortcite{HK} and Schneider et al. (1992)
for a detailed description of its
lensing properties. The area in the lens plane for multiple lensing is
\begin{equation}
\label{SIS-sigma}
\sigma_0 = 16 \pi^3  \frac{\sigma^4_v} {c^4} \left[
\frac{\Dd \Dds}{\Ds}  \right]^2 \, .
\end{equation}
Note that the distance combination which enters equation (\ref{SIS-sigma})
is the same as in the Einstein radius for a point-like gravitational lens.
But, when we consider the projection on the source plane, we get
\begin{equation}
\label{sigma-sis}
\hat \sigma_0 = 
16 \pi^3 \left( \frac{\sigma_v}{c} \right)^4 \Dds^{2} \, .
\end{equation}
As the angular diameter distance between two points at redshifts $z_1$ and
$z_2 > z_1$ is a decreasing function of $\alpha$ \cite{Asada98}, so is the
cross section $\sigma_0$ and, consequently, the optical depth $\tau$.
In Fig.~\ref{fig-IS}a we plot the cross section
$\sigma_0 (z; \alpha)$ relatively to the filled beam case,
evaluated along different line of sghts characterized by different values of the
clumpiness parameter. 

Let us now consider an isothermal sphere with a non zero core radius
$\xi_{\rm c}$. The surface mass density is $ \Sigma(\mathbf {\xi}) =
\frac{\sigma^2_v}{2 G} \frac{1}{\sqrt{\xi^2+\xi_c^2}}$,
which leads to a non constant deflection angle.
The cross section on the lens plane is \cite{HK}
%\begin{displaymath}
\begin{eqnarray}
\label{sigma-is}
\sigma = \left \{
\begin{array}{ll}
\sigma_0 \left[ \left( 1 + 5 \beta - \frac{1}{2} \beta^2 \right) - \frac{1}{2} \sqrt{\beta}
\left( \beta + 4 \right)^{3/2} \right] 
  & \beta< \frac{1}{2}\\
0 & \beta > \frac{1}{2}
\end{array} \right .
\end{eqnarray}
%\end{displaymath}
where $\beta$ is the core radius $\xi_{\rm c}$ in units of the natural length
scale
\begin{equation}
\label{eq-02}
\xi_0 = 4 \pi \left ( \frac{\sigma_v}{c} \right) ^2
\frac{\Dd \Dds}{\Ds}.
\end{equation}
The projected cross section for multiple imaging is, for $\beta < 1/2 $,
\begin{equation}
\widehat{\sigma} = 16 \, \pi^3 C(\beta) \left( \frac{\sigma_v}{c} \right)^4
\Dds^2 \, ,
\end{equation}
where we introduced the quantity $C(\beta) \equiv \sigma / \sigma_0$.
The presence of a core radius lowers the lensing probability, but
does not change its qualitative properties with respect to the
clumpiness parameter, see Fig.~\ref{fig-IS}b.
Note that the
function $C(\beta)$ is not a constant, since $\beta$ is not, with
respect to the clumpiness parameter. The dimensionless core radius
$\beta$ is a function of the redshifts $z, z_s$, and the
cosmological parameters and the clumpiness parameter, via the
length unit $\xi_0$:
\begin{equation}
\beta (z, z_s, \Omega_{\Lambda}, \alpha) = \frac{1}{4\pi} \frac{c^2}{\sigma_v^2}
\xi_c  \, \frac{\Ds}{\Dd \Dds} \, .
\end{equation}

It is interesting to note that, despite the fact that the cross
sections of the considered gravitational lens models differ in the
DR distances ratio, they are both decreasing functions of
$\alpha$. In other words, the monotonic properties of the
cosmological optical depth with respect to $\alpha$ are general.
Finally, it is evident that neglecting $\alpha$ leads, in the
theoretical predictions, to underestimate the probability of
strong lensing, and, in the statistical analysis of high redshift
quasars catalogs, to overestimate the cosmological constant.
This is discussed 
in the next section.

\section{On the expected number of gravitational lenses}

In this section we evaluate numerically the optical depth in a flat
cosmology as a function of the cosmological parameter $\Omega_{\Lambda}$
and of the clumpiness parameter $\alpha$.
In particular we analyze
quantitatively the effect due to neglecting the local inhomogeneity in
calculating the lensing probability in a clumpy universe,
along lines of sight with $\alpha < 1$. 
For this purpose,
we consider the gravitational lenses to be singular isothermal spheres, 
because they allow  a very simple and analytical treatment of the problem,
and have been used in a variety of studies of the statistical properties of the
gravitational lenses (see Chae (2003) and references therein), allowing a direct
comparison of the results.
Moreover, this particular choice does not affect out 
qualitative results.

\begin{figure}
\center{\scalebox{0.45}{\includegraphics{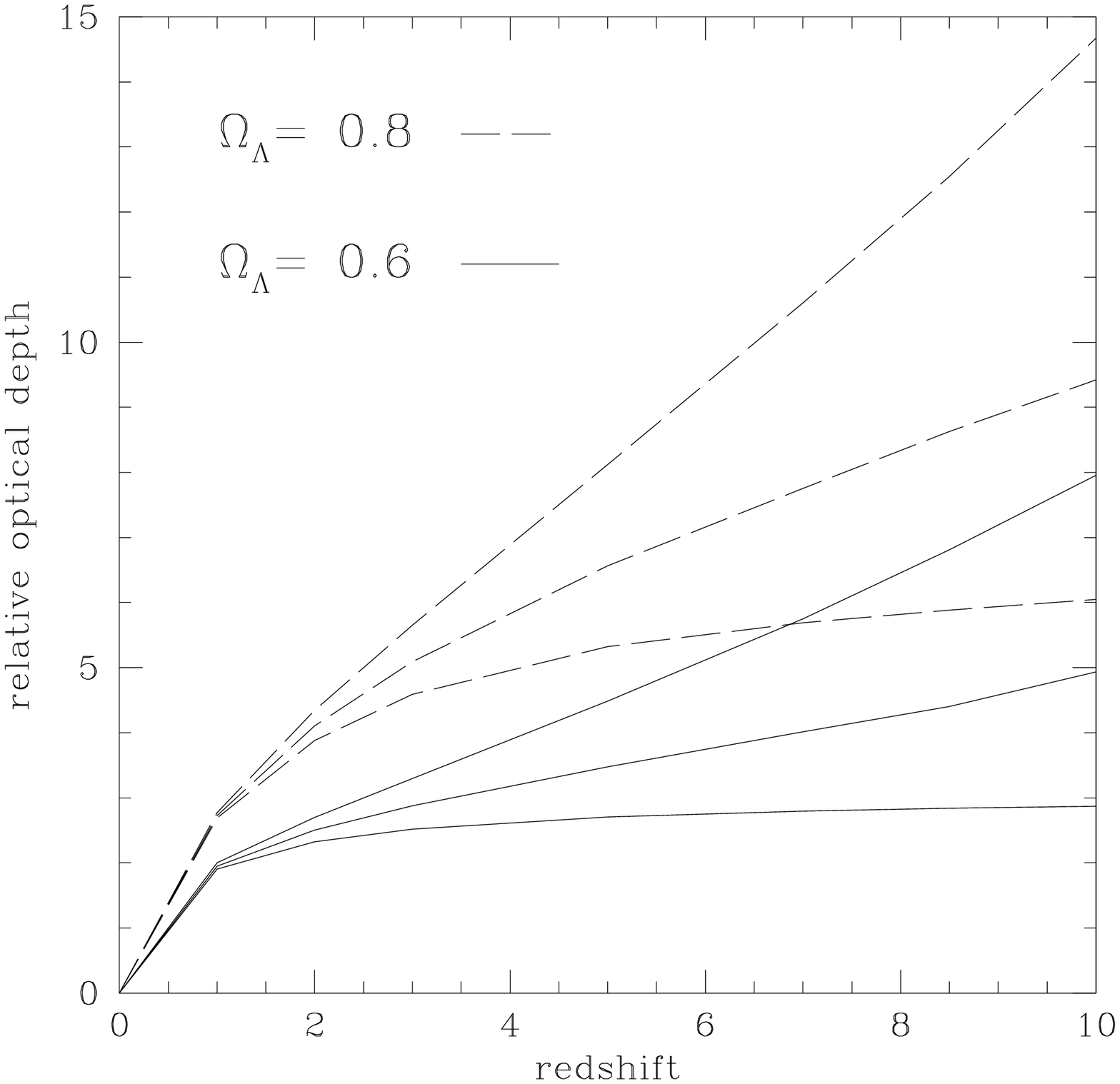}}}
\caption{The optical depth as a function of the redshift for
differ values of $\alpha$ and $\Omega_{\Lambda}$ relatively
to $\tau_0$, the optical depth in the case $\Omega_{\Lambda} =0$
and $\alpha =1$. 
Continuous and dashed curves are for $\Omega_{\Lambda} =
0.6, 0.8$ respectively. The clumpiness parameter is $\alpha = 0,
0.5, 1.0$, from the above curve to the lower one for each value of
$\Omega_{\Lambda}$.
Gravitational lenses are SIS and space-time is
flat. } 
\label{tau-confronto}
\end{figure}

As we are mainly interested in the effects due to a variation of
$\alpha$, we compare the optical depth $\tau (z_s,
\Omega_{\Lambda}, \alpha)$ with $\tau_0$, the numerical value in
the case with vanishing cosmological constant and light 
propagation through filled beams,
so that
our discussion is independent of the numerical parameter 
$ F \equiv 16 \pi^3 \left( \frac{\sigma_v}{c}\right)^4 \left( \frac{c}{H_0}
\right)^3 $ 
which controls the lensing probability. 
In Fig.~\ref{tau-confronto} we plot the relative optical depth $ \tau /
\tau_0$ as a function of the source redshift, for six different
cosmological models, in which $\Omega_{\Lambda} = 0.6, 0.8$, 
considering lines of sight with 
$\alpha  = 0, 0.5, 1$ for each case.
As it is well known,
the strong lensing probability is very sensitive to the value of the
vacuum energy (e.g., Fukugita et al. 1992). 
For any value of $\alpha$, the probability that a source at given
redshift $z_s$ is multiply imaged grows by a factor of $\sim 2$,
if $\Omega_{\Lambda}$ goes from $\sim 0.6$ to $\sim 0.8$. 

\begin{figure*}
\centerline{\scalebox{0.4}{\includegraphics{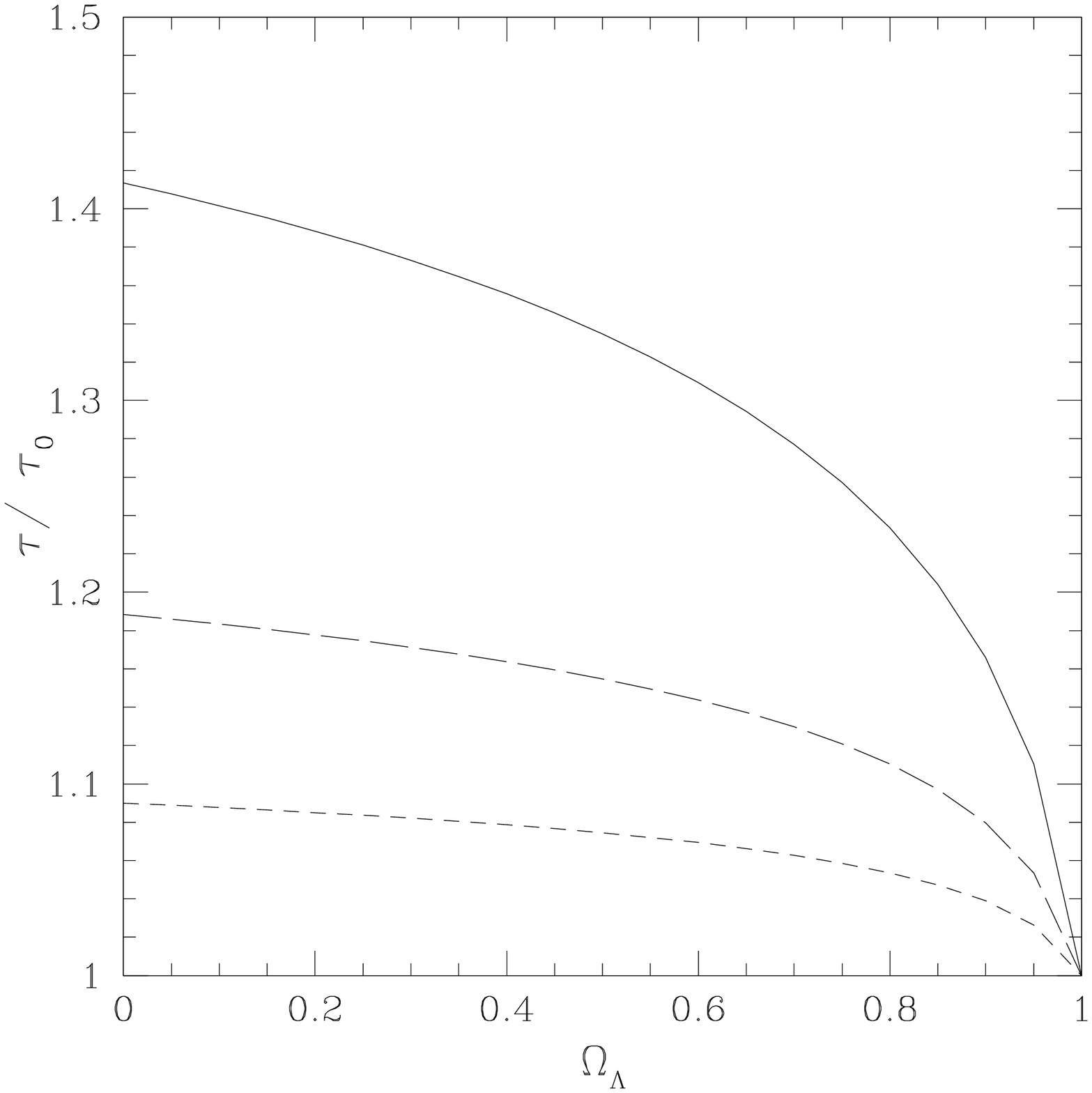}\includegraphics{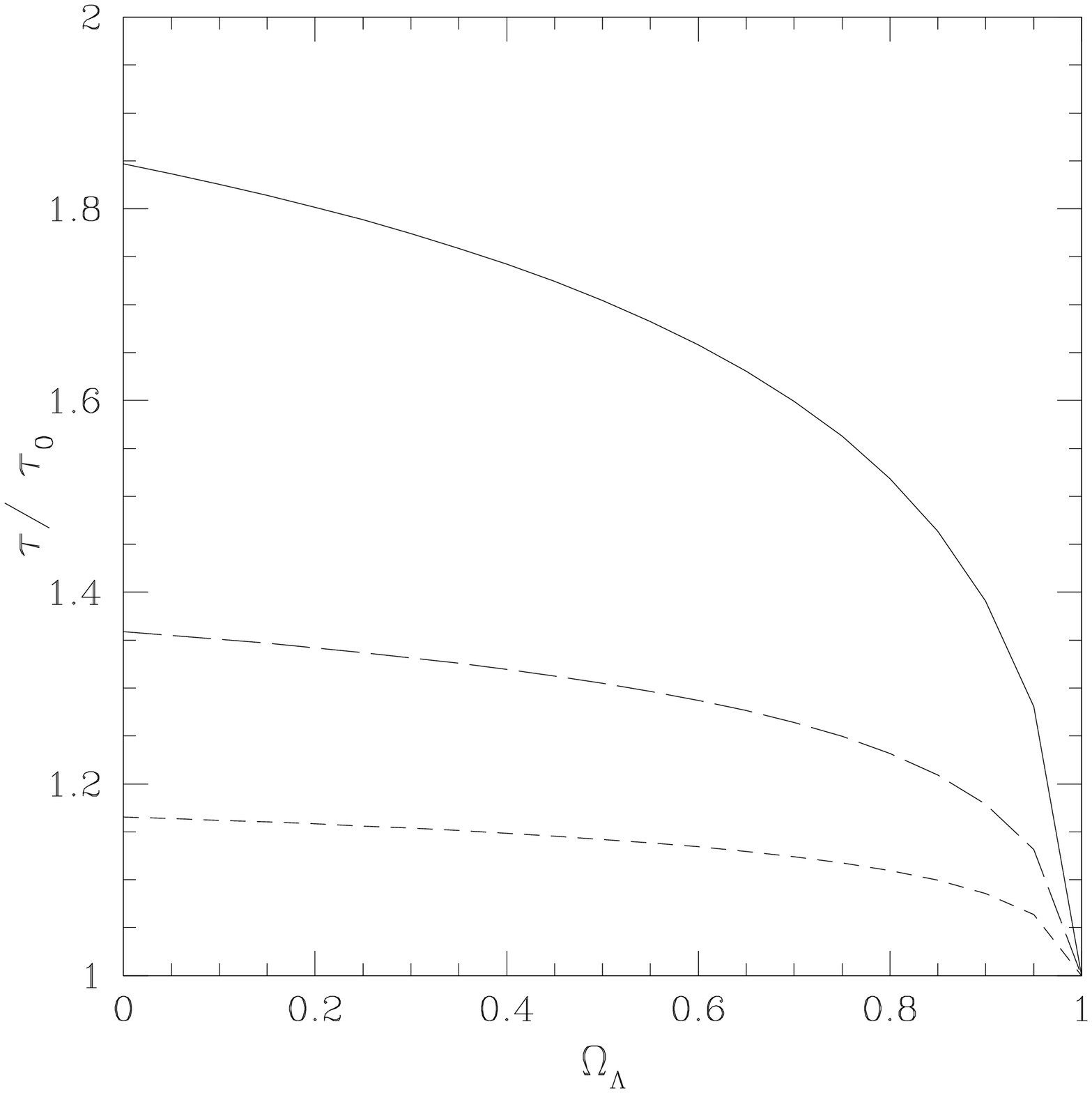}}}
%{\scalebox{0.33}{}
%\centerline{\scalebox{0.33}{\includegraphics{Fig6a.eps}}}
%\centerline{\scalebox{0.33}{\includegraphics{Fig6b.eps}}}
%\centerline{\scalebox{0.33}{\includegraphics{Fig6c.eps}}}
\caption{Relative variation of the optical depth with respect to the 
filled beam case.
The clumpiness parameter is $\alpha = \{ 0, 0.5, 0.75 \}$. Sources are at
redshifts  $z_s=$ 3 (left hand panel), 5 (right hand panel), in a
flat space-time. $\Omega_{\Lambda} = 0.7$}
\label{fig-relative-tau}
\end{figure*}

In order to disentangle the effect of the clumpy distribution of matter 
from that due to the cosmological constant, we consider
$\Delta \tau / \tau_0$,
i.e. the relative variation of the lensing probability  
with respect to the case $\alpha=1$,
 as a function of $\Omega_{\Lambda}$ 
along lines of sight characterized by $\alpha < 1$.
We plot this quantity, evaluated at different source redshifts, 
in  Fig.~\ref{fig-relative-tau}.
As shown above, for any $z_s$ and $\Omega_{\Lambda}$, the lensing
probability is a decreasing function of the clumpiness parameter.
We also notice that the effect of the clumpiness parameter
increases with the redshift of the sources,
if a given direction with $\alpha < 1$ is considered.

The most important
feature to note is that the variation becomes rapidly less
important at larger values of the cosmological constant. The
reason for this effect is twofold. First, for larger values of the
cosmological constant, the influence of all other astrophysical
and cosmological parameters is expected to be less important,
since at high $\Omega_{\Lambda}$, the optical depth is very
sensible to any small change in the value of the cosmological
constant. Second, $\alpha$ enters the equation (\ref{eqn:DR})
as a coefficient of the matter density parameter, determined (in
the flat cosmological models) by the relation $\Omega_{\rm m} = 1 -
\Omega_{\Lambda}$; so it is relatively less important for small
values of the density parameter.
In the most commonly accepted range for the cosmological constant ($0.6
\leq \Omega_{\Lambda} \leq 0.85$), the lensing probability increases
(relatively to a completely homogeneous matter distribution) by a
factor of about $ 7 \%, 17 \%, 30 \% 
$ 
if we consider the clumpiness
parameter $\alpha = 0.75, 0.5, 0$ respectively.

\begin{figure}
\centerline{\scalebox{0.45}{\includegraphics{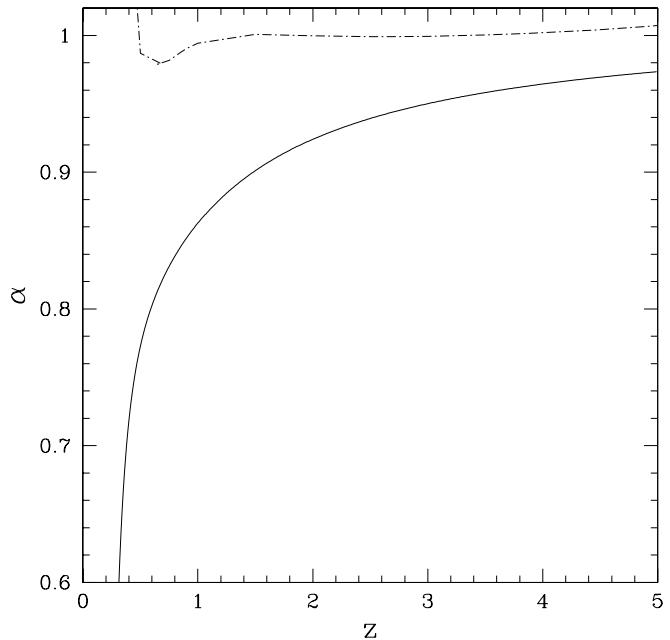}}}
\caption{The peak (continuos line) and the mean value 
(dashed - dotted line) of the clumpiness parameter
in  flat cosmological model with $\Omega_{m} = 0.4$,
calculated via the analytic approximation given in Wang (1999).}
\label{fig:alpha}
\end{figure}

As stated above, 
in order to have a coherent description of the small scale clumpiness
we need to consider the PDF for $\alpha$, since this a direction-dependent
quantity.
The PDF depends on the background FLRW 
cosmological model,
and can be derived from the PDF of the lensing magnification  $\mu (z)$, 
calculated in numerical simulations 
(e.g., Wambsganss et al. 1997, Holz \& Wald 1998, Tomita 1998, M\"ortsell 2002), 
via the relation 
\begin{equation}
\mu (z) =  \left [ \frac{r_1(z)}{r(z, \alpha)} \right ]^2 \, ,
\label{eqn:mu}
\end{equation}
and the DR equation.
As shown by Wang (1999), in general 
the $\alpha$ PDF is peaked at $\alpha < 1$ for any $z$,
but it tends to be more symmetric and 
shows smaller scatter around the peak value as
the redshift increases (see, e.g., Fig.~3 in Wang 1999). 
In Fig.~\ref{fig:alpha} we plot, for a flat spacetime 
with $\Omega_m = 0.4$, both the most likely value
and the average values, as calculated via the 
approximate analytic expressions of the PDF ($\alpha$) given in Wang (1999).

The mean value of ${\alpha}$ is 1 at any redshift in all cosmological
models (if they are FLRW on average as assumed here).
This is indeed the  same proprierty which is known to hold for the PDF
of the lensing magnification $\mu(z)$,
whose basic motivation is the flux conservation (Weinberg 1976).
Therefore we expect that 
the effect of inhomogeneoties on the lensing cross sections 
is reduced in large ensemble of gravitational lenses, or 
very distant sources.

For a gravitational lens at $z \sim 1$, the most likely value is 
$\alpha \sim 0.85$;
this translates in an underestimates of the lensing cross section for a SIS 
(using the fille beam distances and considering a source at 
$z_s \gtrsim 4$) of a factor $\sim 1.15$, see Fig.~\ref{fig-IS}.
These systematic errors decrease for more distant gravitational lenses or 
sources, but 
since the most likely value is always lower than 1,
and in a relatively small sample of gravitational lenses
the mean of a nonsymmetric probability distribution
is not likely the best estimator,
such errors may become not neglibible when evaluating the cosmological
parameters.

\begin{figure}
\centerline{\scalebox{0.45}{\includegraphics{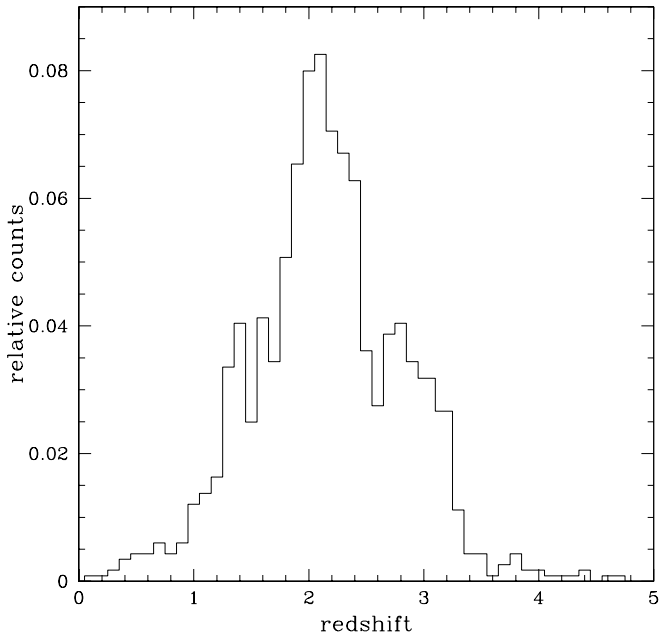}}}
\caption{Redshift distribution of the 1163 luminous
quasars in the catalog used to calculate
the effect of a clumpiness parameter $\alpha \neq 1$. See text for details.}
\label{fig-histogram}
\end{figure}

\begin{figure}
\centerline{\scalebox{0.45}{\includegraphics{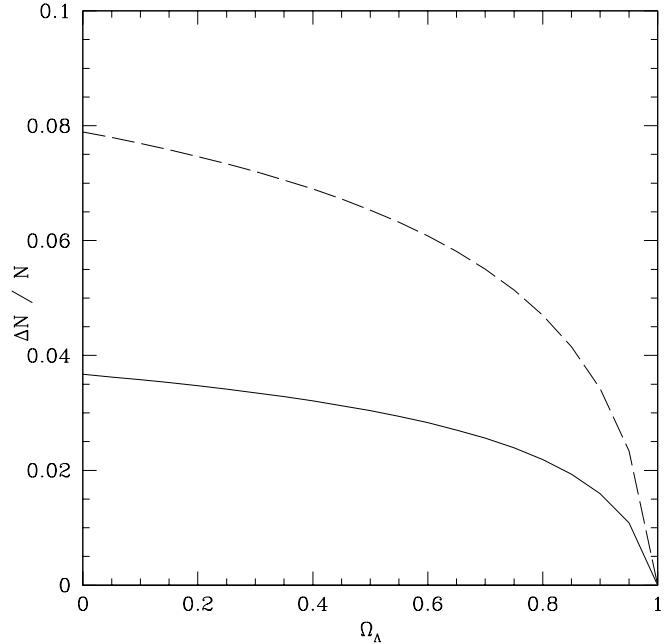}}}
\caption{Relative variation of
the expected number of strong lensed quasars in the given sample
of high luminosity quasars
with respect to the filled beam case, as a function of the cosmological
constant. The clumpiness parameter is $\alpha = 0.75$ (dashed curve)
and $\alpha = 0.95$ (continous curve).
Lenses are modelled as SIS.}
\label{fig-expectedN}
\end{figure}

While a detailed statistical analysis including an accurate determination for
the PDF for any cosmological model is beyond the scope of the present work,
it is anyway important to estimate now an upper limit to the 
possible errors on the predicted number
of gravitational lenses.
We have therefore considered the list of 1163 luminous quasars\footnote{This 
catalog is available at the web address 
{\tt http://vela.astro.ulg.ac.be/themes/extragal/gravlens/bibdat}}. 
which have been observed in the following optical surveys:
CFHT (Crampton, McClure \& Fletcher  1992), 
CFHT (Yee, Filippenko \& Tang 1993),
HST (Maoz et al. 1993),
NOT (Jaunsen et al. 1995).
This catalog contains 7 confirmed gravitational lenses, and its redshift
distribution is plotted in Fig.~\ref{fig-histogram}: 
the peak is at  redshift $z\simeq 2$, and only a small fraction of sources are
beyond $z=3$.
We have calculated the relative variation in the expected number 
of multiply imaged quasars 
considering different values for the clumpiness parameter,
corresponding to peak values of its PDF for $z \simeq 0.5$ and 3.0. 
The lensing galaxies are being modelled as SIS (note that including a 
small core makes the clumpiness effect slighlty larger).
In Fig.~\ref{fig-expectedN}, we plot the relative variation of
the expected number of lenses as a function of $\Omega_{\Lambda}$ for the peak
values $\alpha = 0.75, 0.95$. 
this figure 
shows the upper limit for the systematic errors that can be found 
for the 
predicted numner fo gravitational lenses 
when adopting the simple filled beam hypothesis;
given the property $\bar\alpha = 1$, this could be 
further reduced in upcoming larger surveys.
Note that, since we are considering a flat spacetime
and the effect of the local inhomogeneities increases with $\Omega_{\rm m}$, 
the variation of the
expected number $N$ of multiply imaged quasars is a decreasing function of
the cosmological constant. 
However, only for a very large value of the cosmological constant 
($\Omega_{\Lambda} \ga 0.8$), such a variation becomes rapidly small. 
This makes clear the
point that if in the evaluation of $\tau$ we use angular diameter distances
for a perfectly homogeneous cosmological model, 
we can underestimate the lensing probability and, consequently,
overestimate $\Omega_{\Lambda}$.

\section{Conclusions}

In this work we have investigated whether the
local departures from a completly homogeneous
cosmological model can have observable effects  
in the statistical study of high-$z$ gravitational lenses.
Following the work by
Ehlers \& Schneider \shortcite{ES}, we derived the expressions for
the cosmological optical depth in the framework of a cosmological
model which is FLRW on very large scales (i.e., whose overall
dynamics is very well described by FLRW models) and whose matter
distribution is locally inhomogeneous. 
The direction-dependent clumpiness parameter $\alpha$ 
quantifies the fraction of matter in compact objects
along a given line of sight, and its peak and mean values (as a function fo the
source redshift) are calculated
via the analytical approximation of the PDF given in Wang (1999).
We have paid particular attention to disentangle the 
different role played
by the different notions of distance in the definition of the optical 
depth $\tau (\alpha)$: 
the small scale inhomogeneities along the line of sight do not change the
volume element $dV$, but the strong lensing cross section, 
see equation (\ref{eq:volume}) and Fig.~\ref{fig-IS}.

Up to redshift $z \sim 3$,
the most probable value of the clumpiness parameter is very different 
from the average value 
(which is constrained to be $1$ in a model homogeneous on a
large scale), see Fig.~\ref{fig:alpha},
and the effect may 
be important in statistical analysis of relatively small sets
of gravitationally lensed sources.

Asada (1998) presented a similar calculation 
to the one done here in Sect.~4.
He discussed the influence of
the clumpiness parameter on several gravitational lensing observable
quantities, 
showing that in a clumpy universe deflection angles are smaller and
time delays are longer
than in an homogeneous universe,
given the same lens-source configuration.
However, in contraddiction with Asada's result,
we have found that the gravitational lenses rate is a decreasing function 
of the clumpiness parameter.
We have shown that 
in the empty and filled beam cases different angular distances ratios enter the
optical depth expression (Sect.~3), and
this leads to an higher number of expected  of gravitational lenses
when light beams with $\alpha < 1 $ are considered. 

While a detailed statistical analysis including the effect of small scale
inhomogeneities on the determination of the cosmological constant
is beyond the scope of the present work,
we can already draw some important conclusions and compare our findings with
previous works.
Since the $\alpha$ mean value tends to be 1 
when a sufficiently large number of different 
lines of sight is considered, 
the effect described here tends to be less important 
in large surveys for gravitationally lensed sources.
Moreover, since the peak value also tends to 1 for $z \gtrsim 5 $,
statisical study of very high-redshift sources is less  
affected by the local clumpiness along the different lines of sight.
On small set of lensed sources, 
since the peak value of the clumpiness parameter is always lower than 1, 
using the filled beam angular diameter distances leads 
to an overestimate of the number of the expect gravitational lenses (Fig.~9),
and, consequently, to overestimate the cosmological cosntant.

It is interesting to note that the same qualitative result
(on the determination of the cosmological constant) 
has been found in previous  works in which the authors have taken into accounts
the effect on the distance-redshift relation playd by
the matter distribution inhomogeneities.
Kantowski (1998) described the effects of inhomogeneities on the
determination of the cosmological constant and 
$\Omega_m$ using a ``Swiss cheese'' model to derive analytic expressions for 
the distance-redshift relation. 
He has shown that, when analyzing high-$z$ standard candles, assuming a
completely homogeneous matter distribution leads to   
overestimating  the cosmological constant if the filled beam hypotheisis is
always used (see, e.g., Fig.~8 in Kantowski (1998)).

In a recent work,
Barris et al. (2004) analyze a set of 194 Type Ia SNe 
at $z>0.01$ (including the 23 discovered in the IfA Deep Survey
in the range $z = 0.34- 1.03$)
using both the filled beam and the empty beam assumption 
for {\em all} the lines of sight
to calculate luminosity
distances\footnote{Note, however, that using the empty beam 
approximation for all the
lines of sight is not a coherent description, since 
in an inhomogeneous Universe not all the light beams can be devoided of matter,
and more realtic distribution for $\alpha$ has to be chosen
to describe the clumpiness effects using the DR distances.}
.
Although the presence of small
scale inhomogeneities is far from eliminating the need of a vacuum energy
contribution to explain the new data, they do change the final estimate of
the cosmological parameters 
$\Omega_{\rm m}$ and $\Omega_\Lambda$: 
indeed, when using the angular-diameter distances with $\alpha=0$,
the confidence contours for the comsological parameters 
are shifted to lower values of $\Omega_\Lambda$ 
and higher values of $\Omega_{\rm m}$

The same effect has been found by authors 
which investigated the effect of large
scale inhomogeneity in the matter distribution.
Tomita (2001) considered a local void 
on scales of 200-300 Mpc around our Galaxy to interpret the 
high-$z$ SN data, effectively describing it with 
with $\alpha < 1$ in the DR distances.
He finds that the data (available at the time)
coule be well fitted with $\Omega_{\Lambda} \sim 0.4$
(value estimated in the overdense outer region).
It is certainly interesting to test these findings with the more recent and
larger set of data available at the moment.
In conclusion, a precise determination of the cosmological parameters 
using both the
Hubble diagram of the high-$z$ Type Ia SN and the statistical properties of
gravitational lenses requires an accurate determination of the effect of the
local clumpiness in the matter distribution on the light propagation.

\section*{Acknowledgments}

GC wishes to thank N.D. Morgan, N.R. Napolitano 
and Y. Wang for stimulating discussions,
and the referre for a careful reading 
and insightful remarks.
 Special thanks to A. Pospieszalska-Surdej
 and the Gravitational Lenses group at the
 Institut d'Astrophysique et de G\'eophysique
 (University de Li\`ege) 
 for providing to the community the catalog of the quasars 
 searched for multiple lensed images in optical surveys.
GC is supported by the Euro3D RTN postdoctoral fellowship.

This work is dedicated to the beloved 
memory of Professor Ruggiero de Ritis.


\begin{thebibliography}{99}

\bibitem[\protect\citename{Asada }1998]{Asada98}
Asada H., 1998, ApJ, 501, 473

\bibitem[\protect\citename{Barris et al. }2004]{Barris}
Barris B.J. et al., 2003, ApJ, 602, 571

\bibitem[\protect\citename{Buchert }2000]{Buchert}
Buchert T., 2000, 
in: 9th JGRG Meeting,
Hiroshima 1999, Y. Eriguchi et al. (eds.), 2000, p. 306
(preprint gr-qc/0001056)

\bibitem[\protect\citename{Chae }2003]{Chae2003}
Chae K.-H., 2003, MNRAS, 346, 746
 
\bibitem[\protect\citename{Cheng \& Krauss }2000]{Cheng}
Cheng N.Y., Krauss L.M., 2000, Int. J. Mod. Ph., A 15, 697

\bibitem[\protect\citename{Chiba \& Yoshii }1999]{Chiba99}
Chiba M., Yoshii Y., 1999, ApJ, 510, 42

\bibitem[\protect\citename{Claeskens \& Surdej }2001]{claeskens}
Claeskens J.F., Surdej J., 2002 A\&AR, 10, 263

\bibitem[\protect\citename{}]{Crampton}
Crampton D., McClure R.D., Fletcher J. M., 1992, ApJ, 392, 23

\bibitem[\protect\citename{Dyer \& Oattes }1988]{DO88}
Dyer C.C., Oattes, L., 1988, ApJ, 326, 50

\bibitem[\protect\citename{Dyer \& Roeder }1972]{DR72}
Dyer C.C., Roeder R.C., 1972, ApJ, 174, L115

\bibitem[\protect\citename{Dyer \& Roeder }1973]{DR}
Dyer C.C., Roeder R.C., 1973, ApJ, 180, L31

\bibitem[\protect\citename{Ehlers et al. }1968]{Ehlers68}
Ehlers J., Geren P., Sachs R.K., 1968, J. Math. Phys., 9, 1344

\bibitem[\protect\citename{Ehlers \& Schneider }1986]{ES}
Ehlers J., Schneider P., 1986, A\&A, 168, 57

\bibitem[\protect\citename{Ellis }1995]{Ellis95}
Ellis, G.F.R. 1995, {\em Observations and cosmological models} in
{\em Galaxies and the Young Universe}, eds. H. von Hippelein, K.
Meisenheimer \& J.H. Rose, Springer, Berlin

\bibitem[\protect\citename{Fukugita \& Turner }1991]{Fukugita91}
Fukugita M., Turner E.L., 1991, MNRAS, 253, 99

\bibitem[\protect\citename{Fukugita et al. }1992]{Fukugita92}
Fukugita M., Futamase T., Kasai M., Turner E.L., 1992, ApJ, 393, 3

\bibitem[\protect\citename{Hamana et al. }1997]{Hamana97}
Hamana T., Futamase K., Futamase T., Kasai M., 1997, MNRAS, 287, 341

\bibitem[\protect\citename{Hinshaw \& Krauss }1987]{HK}
Hinshaw G., Krauss L.M., 1987, ApJ,  320, 468

\bibitem[\protect\citename{Holz }1998]{Holz1998}
Holz D.E., 1998, ApJ, 506, L1

\bibitem[\protect\citename{Holz }1998]{Holz1998}
Holz D.E., Wald R.M. 1998, Phys. Rev. D 58, 063501

\bibitem[\protect\citename{}]{Jaunsen}
Jaunsen A. O., Jablonski M., Pettersen B.R., Stabell R., 1995, A\&A, 300, 323

\bibitem[\protect\citename{Kantowski }1998]{kanto1998}
Kantowski R., 1998, ApJ 507, 483

\bibitem[\protect\citename{Kantowski et al. }2000]{kanto2000}
Kantowski R., Kao J.K, Thomas R.C., 2000, ApJ, 545, 549

\bibitem[\protect\citename{Kayser et al. }1997]{kayser97}
Kayser R., Helbig P., Schramm T., 1997, A\&A, 318, 680

\bibitem[\protect\citename{Kochaneck }1993]{Kochanek93}
Kochanek C.S., 1993, ApJ, 419, 12

\bibitem[\protect\citename{Kochaneck }1996]{Kochanek96}
Kochanek C.S., 1996, ApJ, 466, 638

\bibitem[\protect\citename{Kochaneck et al. }1999]{Kochanek99}
Kochanek C.S. et al. in Holt S.S., Smith E.P., eds,
AIP Conf. Proc. Vol. 470
th Astrophys. Conf.,
After the Dark Ages: When Galaxies Were Young
(The Universe at $2 < z < 5$). Am Inst. Phys., New York, p. 163

\bibitem[\protect\citename{Krasi\'nsky}1997]{kra97}
Krasi\'nski A., 1997, Inhomogeneous Cosmological Model, Cambridge
University Press

\bibitem[\protect\citename{Kristian \& Sachs }1965]{sachs}
Kristian J., Sachs R.K., 1965, ApJ, 143, 379

\bibitem[\protect\citename{Kuhlen et al. }1988]{kuhlen}
Kuhlen M., Keeton C.R., Madau P., 2003, ApJ, 601, 104

\bibitem[\protect\citename{Linder }1988]{linder1988}
Linder E.V., 1988, ApJ, 497, 28

\bibitem[\protect\citename{Mao }1991]{Mao91} 
Mao S., 1991, ApJ, 380, 9

\bibitem[\protect\citename{Maoz \& Rix }1993]{maoz93}
Maoz D., Rix H.-W., 1993, ApJ, 416, 425

\bibitem[\protect\citename{Maoz et al. }1993]{maoz2}
Maoz D. et al., 1993, ApJ, 409, 28

\bibitem[\protect\citename{Morgan }2002]{morganPhd}
Morgan N. D., 2002, PhD thesis, Massachusetts Institute of Technology

\bibitem[\protect\citename{Morgan et al. }2004]{morgan04}
Morgan N. D., Caldwell  J.A.R., Schechter P.L., Dressler A., 
Egami E.; Rix  H.-W., 2004, AJ, 127, 2617

\bibitem[\protect\citename{Mortsell }2002]{}
M{\"o}rtsell E., 2002, AA 382, 787

\bibitem[\protect\citename{Perlmutter et al. } 1999]{perlmutter}
Perlmutter S. et al., 1999, ApJ, 517, 565

\bibitem[\protect\citename{Pyne \& Birkinshaw }2004]{pyne}
Pyne T., Birkinshaw M., 2004, MNRAS, 348, 581

\bibitem[\protect\citename{Rusin et al. }2002]{rusin2002}
Rusin D., Norbury M., Biggs A.D. et al., 2002, MNRAS, 330, 205 

\bibitem[\protect\citename{Russ et al. }1997]{russ97}
Russ H. et al., 1997, PRD, 56, 2044

\bibitem[\protect\citename{Schmidt et al. }1998]{schmidt}
Schmidt B.P. et al., 1998, ApJ, 507, 46

\bibitem[\protect\citename{Schneider, Ehlers \& Falco }Schneider et al. 1992]{SEF}
Schneider P., Ehlers J., Falco E.E., 1992, Gravitational Lenses,
Springer-Verlag, Berlin

\bibitem[\protect\citename{Seitz et al. }1994]{seitz}
Seitz S., Schneider P., Ehlers J., 1994, Class. Quant. Grav., 11, 2345

\bibitem[\protect\citename{Sereno et al. }2001]{}
Sereno M., Covone G., Piedipalumbo E., de Ritis R., 2000, MNRAS, 327, 517

\bibitem[\protect\citename{Sereno et al. }2002]{}
Sereno M., Piedipalumbo E., Sazhin M.V., 2002, MNRAS, 335, 1061

\bibitem[\protect\citename{Stoeger et al. }1995]{Stoeger}
Stoeger W.R, Maartens R., Ellis G.F.R., 1995, ApJ, 443, 1

\bibitem[\protect\citename{Tomita }1998]{Tomita}
Tomita K., 1998, Prog. Theor. Phys., 100, 79

\bibitem[\protect\citename{Tomita }2001]{Tomita01}
Tomita K., 2001, MNRAS, 326, 287

\bibitem[\protect\citename{Turner, Ostriker \& Gott }1984]{turner}
Turner E.L., Ostriker J.P., Gott J.R., 1984, ApJ, 284, 1

\bibitem[\protect\citename{Wambsganss at al. }1997]{wamb97}
Wambsganss J., Cen R., Xu G., Ostriker J.P. 1997, ApJ 475, L81

\bibitem[\protect\citename{Wang at al. }2000]{wang}
Wang L., Caldwell R.R., Ostriker J.P., Steinhardt P.J., 2000, ApJ, 530, 17

\bibitem[\protect\citename{}]{}
Wang Y., 1999, ApJ, 525, 651  

\bibitem[\protect\citename{}]{}
Wang Y., 2000, ApJ, 536, 531  

\bibitem[\protect\citename{}]{}
Wang Y., Holz D.E., Munshi D., 2002, ApJ 572, L15

\bibitem[\protect\citename{}]{}
Weinberg S., 1976, ApJ 208, L1

\bibitem[\protect\citename{}]{yee}
Yee H. K. C., Filippenko A.V., Tang D., 1993, AJ, 105, 7

\bibitem[\protect\citename{Zel'dovich }1964]{Zeldovich}
Zel'dovich Y.B., 1964, SvA, 8, 13


\end{thebibliography}
\end{document}